\documentstyle[aps,preprint,amsfonts]{revtex}
 \tighten
\begin{document}
\draft
\title{Self-Organized Critical Models Without Local Particle Conservation Laws
 On Superlattices}
\author{H. F. Chau}
\address{
 School of Natural Sciences, Institute for Advanced Study, Olden Lane,
 Princeton, NJ 08540, U.S.A.
}
\date{March 7, 1995}
\preprint{IASSNS-HEP-95/17}
\maketitle
\begin{abstract}
 We consider simple examples of self-organized critical systems on
 one-dimensional superlattices without local particle conservation laws. The
 set of all recurrence states are also found in these examples using a
 method similar to the burning algorithm.
\end{abstract}
\section{Introduction}
 In 1987, Bak {\em et al.} proposed an interesting idea called self-organized
 criticality, suggesting that many physical systems could evolve under their
 own dynamics without any fine tuning parameters to states without any
 characteristic time and length scales \cite{BTW}. They illustrated the idea
 using a simple cellular automaton model. Later on, their automaton model is
 proved to be commutative in the sense that the order of ``particle addition
 operations'' does not affect the outcome of the final state \cite{ASM}. Now,
 this model is widely known as the Abelian Sandpile Model (ASM). In addition to
 the cellular automaton models, various kinds of continuous and
 lattice-continuous models have shown to exhibit self-organized criticality.
\par
 In particular, the self-organized critical model of dissipative transport
 using stochastic partial differential equation \cite{Hwa} has received much
 attention recently. This model proves that the existence of a local
 (particle) conservation law is not a necessary condition for the exhibition of
 self-organized criticality. Further discussions on the importance of particle
 conservation law can be found elsewhere \cite{Spec}. Furthermore, the
 existence of local particle conservation law is not an essential feature for
 cellular automaton models (i.e. models on finite grid points and each grid
 point allows only a finite number of states) as well. The well-known forest
 fire model is an example of this kind \cite{FFM}.
\par
 In this paper, we study two new examples of Abelian Sandpile Models on
 one-dimensional superlattices --- the simplest possible kind of
 self-organized critical model without local particle conservation laws. The
 examples we provide here can be generalized easily to higher dimensions
 although exact calculation of the scaling exponents and the recurrence phase
 space configurations is much more difficult. Studies of similar
 one-dimensional models can be found elsewhere \cite{Other}. A useful method,
 inspired by the idea of the burning algorithm \cite{BA}, to find the
 recurrence phase space which constitute an important step in showing the
 criticality of our model is also introduced. This method works well for
 systems of low spatial dimensions.
\section{Example 1}
 Consider a collection of $2N$ sites labeled from 1 to $2N$ and we associate
 an integer $h_i$ called the local height to each site. Using the same rules
 for the Abelian Sandpile, sites with local height greater than 0 is said to
 be unstable and particle redistribution occurs as follows: (a) for any
 odd-numbered site $2i\!+\!1$, two particles will be lost in the next timestep.
 Each of its two nearest neighbors, i.e. sites $2i$ and $2i\!+\!2$, receives
 two particles in the next timestep. Clearly two extra particles are created in
 the particle redistribution (or toppling) process. (b) for any even-numbered
 site $2i$, four particles will be lost in the next timestep. Each of its two
 nearest neighbors, i.e. sites $2i\!-\!1$ and $2i\!+\!1$, receives a single
 particle. Thus, two particles are dissipated in the process. The odd and
 even-numbered sites are called creative and dissipative respectively. Open
 boundary conditions (i.e. particles are allowed to flow out of the system from
 both ends) are used. Therefore, the toppling matrix \cite{ASM} of the system
 is given by
\begin{equation}
 \Delta \equiv \Delta^{(2N)} = \left[
 \begin{array}{rrrrrrrr} 2 & -2 & & & & & & \\
 -1 & 4 & -1 & & & & & \\
 & -2 & 2 & -2 & & & & \\
 & & -1 & 4 & -1 & & & \\
 & & & \cdots & \cdots & \cdots & & \\
 & & & & \cdots & \cdots & \cdots & \\
 \\ \end{array} \right] \label{E:Delta1_Def}
\end{equation}
 where the superscript $(2N)$ denotes the total number of sites in the system
 and hence the size of the matrix. Clearly we have constructed a
 one-dimensional superlattice model with equal number of dissipative and
 creative sites.
\par
 First we find the total number of recurrence states on the system which
 equals $\det\Delta$ \cite{ASM}. By some elementary row and column
 transformations, it is easy to show that
\begin{equation}
 \det \Delta^{(2k)} = 4 \det \Delta^{(2k-2)} - 4 \det \Delta^{(2k-4)}
 \label{E:Recur1}
\end{equation}
 for $k\geq 3$. Since $\det\Delta^{(2)} = 6$ and $\det\Delta^{(4)} = 20$,
 we conclude that
\begin{equation}
 \det \Delta^{(2N)} = (2N+1)\ 2^N \mbox{.} \label{E:DDelta1}
\end{equation}
\par \indent
 Now we want to find all the $(2N+1)2^N$ recurrence states of the system using
 an idea inspired by the burning algorithm \cite{BA}. We represent system
 configurations by row vectors of length $2N$. It has been shown that a system
 configuration $\alpha \equiv (\alpha_1,\ldots ,\alpha_{2N})$ is in the set of
 all recurrence states $\Omega$ if and only if we can find an unstable state
 $\beta = (\beta_1,\ldots ,\beta_{2N})$ which topples to $\alpha$ with all the
 sites topple at least once in the process \cite{Speer,GASM}. Clearly,
 $\beta_i = \alpha_i + \sum_j n_j \Delta_{ji}$ for some $n_j \in {\Bbb Z}^+$.
 To test if a particular state $\alpha$ is recurrence, we choose a set of $n_i
 \in {\Bbb Z}^+$ with $a_i = \sum_j n_j \Delta_{ji} \geq 0$ for all $i$. The
 existence of such set of $n_i$ has been proved in Proposition~1 of
 Reference~\cite{Aval_Finite}. Nevertheless the choice of $n_i$ is not unique.
 We add $a_i$ particles to site $i$ for all $i$ at the same time when the
 system is in configuration $\alpha$. Then the resultant configuration after
 toppling equals $\alpha$ if and only if $\alpha$ is a recurrence state
 \cite{GASM,Inv_A}.
\par
 In the present case, we choose $n_i = 1$ for all $i$, which is equivalent
 to adding a single particle to sites $1$ and two particles to site $2N$ at the
 same time. Since at most $\Delta_{ii}$ particles are removed from site $i$
 whenever it is unstable each time, the possible local heights an odd (even)
 site can be when the system reaches its recurrence phase space are -1 and 0
 (-3, -2, -1 and 0). An odd (even) site is called ``absorbing'' if and only if
 its local height equals -1 (-2 or -3). Obviously, these are sites which can
 ``absorb'' the particles coming from their neighbors during toppling for
 exactly one time.
\par \medskip \noindent
Claim: {\em any system configuration with more than one absorbing site is not
 a recurrence configuration.}
\par \noindent
Proof: Suppose $\alpha$ has more than one absorbing site, and
 the left most and right most absorbing sites are denoted by $l$ and $r$
 respectively (i.e., site $i$ is not absorbing if $i < l$ or $i > r$). Upon
 addition of a particle to site $1$ and two particles to site $2N$, it is easy
 to verify that exactly one toppling will occur in sites $1,2,\ldots ,l-2,
 l-1,r+1,r+2,\ldots ,2N$. After that, the avalanche stops because both $l$
 and $r$ ``absorb'' the incoming particles and prevent further toppling. Since
 $l < r$ (or else $\alpha$ has only one absorbing site), the system does not
 return to $\alpha$ after the avalanche. Thus $\alpha\not\in\Omega$.
\hfill $\Box$
\par \medskip
 The remaining possible recurrence state configurations are those $\alpha$
 with at most one absorbing site. The total number of such states, $T$, is
 given by:
\begin{equation}
 T = 2^N + \sum_{i=1}^{N} 2^N + \sum_{i=1}^{N} 2\ 2^{N-1} = (2N+1)\ 2^N
\end{equation}
 where the first term is the total number of stable configurations without any
 absorbing site, the second (the third) terms are the numbers of possible
 recurrence configurations with exactly one odd and no even (one even and no
 odd) absorbing site. Since $T = \det\Delta$, we conclude that stable
 configurations with at most one absorbing site are the only elements in the
 recurrence phase space $\Omega$. The above method of finding recurrence
 configurations is effective whenever the spatial dimension of the system is
 low.
\par
 Having explicitly find out all the elements in $\Omega$, we can proceed to
 show that the system is indeed self-organized critical. We define the
 avalanche size $s$ to be the total number of toppling occur during an
 avalanche (i.e., sites with multiple toppling are counted multiple number of
 times). Direct calculation tells us that if we add a single particle to site
 $i$ on a recurrence system configuration without any absorbing site,
\begin{mathletters}
\begin{equation}
 s = \left\{ \begin{array}{cl} i(2N+1-i) & \mbox{if~} h_i = 0 \\
 0 & \mbox{otherwise} \end{array} \right. \mbox{.}
\end{equation}
 Similarly, if the particle is added to a recurrence configuration with an
 absorbing site at $k$,
\begin{equation}
 s = \left\{ \begin{array}{cl} 0 & \mbox{if~} i = k \mbox{~or~} h_i < 0 \\
 (2N+1-i)(i-k) & \mbox{if~} i > k \mbox{~and~} h_i = 0 \\
 i(k-i) & \mbox{if~} i < k \mbox{~and~} h_i = 0 \end{array} \right. \mbox{.}
\end{equation}
\label{E:D1s}
\end{mathletters}
So under a uniform particle addition and in the $N\!\rightarrow\!\infty$
 limit, the distribution of avalanche size can be well approximated by
\begin{equation}
 D (s) = \left\{ \begin{array}{cl} \frac{1}{4} & \mbox{if~} s = 0
 \vspace{0.1in} \\ \frac{3}{4} \ \frac{2}{N^2} \left[ 1 - \frac{s}{N^2} \right]
 & \mbox{if~} 0 < s < N^2 \vspace{0.1in} \\ 0 & \mbox{otherwise} \end{array}
 \right. \mbox{.} \label{E:Distribution1}
\end{equation}
 Thus if we Fourier transform $D (s)$, $1/f^2$ scaling in avalanche size $s$ is
 observed. Therefore this model is self-organized critical although its scaling
 exponent is trivial.
\par
 To some extend, the above example does not completely demonstrate that
 particle conversation law is not a necessary condition for the exhibition of
 self-organized criticality in cellular automaton models. If we rescale all the
 even-number sites by $h_{2i}\!\longrightarrow\!h_{2i} / 2$, then the toppling
 matrix becomes that of the one-dimensional Abelian Sandpile Model. In fact,
 this example is equivalent to the one-dimensional ASM which allows
 half-integral local heights for all the even-numbered sites. Besides, the
 even-numbered sites will receive only one half of a unit of particle each time
 when something is dropped onto them. The distribution of avalanche size $D(x)$
 can be calculated using the piecewise linear relationship found by Chau and Ho
 \cite{Relative_Scale} and it turns out to be the same as those given by
 Eqs.~(\ref{E:D1s}) and~(\ref{E:Distribution1}).
\section{Example 2}
 We now provide another example on one-dimensional superlattice which is not
 equivalent to any ``ordinary'' one-dimensional sandpile models, whose toppling
 rules are translational invariant at all sites except possibly at system
 boundaries, under local rescaling. More precisely, this model is not
 equivalent to any one-dimensional ASM whose toppling matrix is of Toeplitz
 form. Again, we apply the rules of Abelian Sandpile to a collection of $2N$
 sites labeled from 1 to $2N$. The toppling rules are given by: (a) for any
 odd-numbered site $2i\!+\!1$, it will lose four particles in the next
 timestep. Two of them are delivered to site $2i\!+\!3$, and one of them to
 site $2i\!+\!4$ while the remaining one is dissipated in the process. Thus,
 odd-numbered sites are dissipative. (b) for any even-numbered site $2i$, it
 will lose two particles in the next timestep. Two of them are transported to
 site $2i\!-\!3$, one of them to site $2i\!-\!2$. Thus, even-numbered sites
 are creative. The toppling matrix of the system is given by
\begin{equation}
 \Delta \equiv \Delta^{(2N)} = \left[
  \begin{array}{rrrrrrrr} 4 & 0 & -2 & -1 & & & \\
   0 & 2 & & & & & & \\
   & & 4 & 0 & -2 & -1 & & \\
   -2 & -1 & 0 & 2 & & & & \\
   & & & & \cdots & \cdots & \cdots & \cdots \\
   & & \cdots & \cdots & \cdots & \cdots & & \\
   \\ \end{array} \right] \mbox{.} \label{E:Delta2_Def}
\end{equation}
\indent
 Using similar idea as in Example~1, it is easy to show that
\begin{equation}
 \det\Delta^{(2N)} = \left( N+1 \right) 4^N \mbox{.} \label{E:DDelta2}
\end{equation}
 We use the same technique before to find all the $(N+1)4^N$ recurrence states
 of the system. In this case, we choose $n_i = 1$ for all $i$, which is
 equivalent to adding four particles to sites 1 and $2N\!-\!1$, together with
 two particles to sites $2$ and $2N$ all at the same time. Since at most
 $\Delta_{ii}$ particles are removed from site $i$ whenever it becomes unstable
 each time, the possible states an even (odd) site can be when the system
 reaches the recurrence phase space are -1 and 0 (-3, -2, -1 and 0). For any
 stable system configuration $\alpha \equiv (h_i)$, we define
\begin{mathletters}
\begin{equation}
 u_{min} = \left\{ \begin{array}{cl} i & \mbox{if~} h_1,h_3,
 \ldots ,h_{2i-3} \geq -1 \mbox{~and~} h_{2i-1} < -1 \\
 N\!\!+\!\!1 & \mbox{if~} h_1,h_3,\ldots ,h_{2N-1} \geq -1
 \end{array} \right. \mbox{,} \label{E:Def_u_min}
\end{equation}
 and
\begin{equation}
 l_{max} = \left\{ \begin{array}{cl} i & \mbox{if~} h_{2i+2},
 h_{2i+4},\ldots ,h_{2N} = 0 \mbox{~and~} h_{2i} = -1 \\
 0 & \mbox{if~} h_2,h_4,\ldots ,h_{2N} = 0
 \end{array} \right. \mbox{.} \label{E:Def_l_max}
\end{equation}
\end{mathletters}
For the odd-numbered sites, exactly two particles are received per toppling
 in the particles flows into them; while for the even-numbered sites, the only
 one particle is received per toppling. So odd-numbered sites with local height
 equals -3 or -2 (or even-numbered sites with local height equals -1) can
 ``absorb'' the particles coming from their neighbors during toppling for
 exactly once; while odd-numbered sites with local height equals -1 or 0 (or
 even-numbered sites with local height equals 0) becomes unstable whenever
 particles flows into them during an avalanche. Thus $2u_{min}\!-\!1$ is the
 minimum odd-numbered site (while $2l_{max}$ is the maximum even-numbered site)
 in our finite superlattice which can ``absorb the stress'' whenever particles
 topple onto it for exactly one time during an avalanche.
\par \medskip \noindent
Claim: {\em any system configuration with $u_{min} < l_{max}$ is not a
 recurrence state.}
\par \noindent
Proof: Consider a system configuration $\alpha$ with $u_{min} <
 l_{max}$. Upon addition of 2 particles to sites 1 and $2N\!-\!1$ and 1
 particle to sites 2 and $2N$, it is easy to verify that exactly one toppling
 will occur in the following sites $1,2,\ldots ,2u_{min}\!-\!2$, $2u_{min}$,
 $2l_{max}\!-\!1$, $2l_{max}\!+\!1,2l_{max}\!+\!2,\ldots ,2N$. After that, the
 avalanche stops because both sites $2u_{min}\!-\!1$ and $2l_{max}$ ``absorb''
 the incoming particles and prevent further toppling. Since $u_{min} <
 l_{max}$, sites such as $2u_{min}\!-\!1$ and $2l_{max}$ will not topple during
 the avalanche and hence after the system will not relax back to $\alpha$. Thus
 $\alpha\not\in\Omega$.
\hfill $\Box$
\par \medskip
 Thus the remaining possible recurrence state configurations are those $\alpha
 \equiv (h_i )$ with $-3\leq h_{2i-1}\leq 0$ and $-1\leq h_{2i}\leq 0$ for
 $i=1,2,\ldots ,N$, and with $u_{min}\geq l_{max}$. The total number of such
 states, $T$, is given by:
\begin{equation}
 T = \sum_{i=1}^{N} \sum_{j=1}^{i} 2^{i-1}2\ 4^{N-i}2^{j-1} + \sum_{i=1}^{N}
 2^{i-1}2\ 4^{N-i} + 2^N 2^N
\end{equation}
 where the first term is the number of configurations with $u_{min}\leq N$ and
 $l_{max}\geq 1$, second term is the number of configurations with $l_{max}=0$,
 and third term is the number of configurations with $u_{min}=N\!+\!1$. After
 some computation, we find $T=(N\!+\!1)4^N=\det\Delta$, and hence the set of
 all recurrence states of the system is
\begin{eqnarray}
 \Omega & = & \{ \alpha = (\alpha_i ) : \alpha_{2j-1}\in\{-3,-2,-1,0\},
 \alpha_{2j}\in\{-1,0\} \nonumber \\ & & \mbox{~for~} i=1,2,\ldots ,N
 \mbox{~and~} u_{min} (\alpha) \geq l_{min} (\alpha) \} \mbox{.}
 \label{E:Char_Omega2}
\end{eqnarray}
\indent
 Now, we go on to show that the system is indeed self-organized critical.
 Unlike Example~1, it is not easy to argue the distribution of avalanche size
 $D(s)$ owing to the complexity of the recurrence phase space $\Omega$. So we
 take the alternative approach by calculating the two point correlation
 function $G_{ij}$ of the system which is defined as the average number of
 toppling occurs in site $j$ given that a particle is introduced to site $i$,
 and is given by $G_{ij} = \Delta_{ij}^{-1}$ \cite{ASM}.
\par
 In the Appendix, we show that
\begin{mathletters}
\begin{equation}
 G_{2i,2j} = \left\{ \begin{array}{cl} \frac{j(N-i)}{4(N+1)} & \hspace{0.2in}
 \mbox{if~} i > j \vspace{0.08in} \\ \frac{(i+1)(N+1-j)}{4(N+1)} &
 \hspace{0.2in} \mbox{otherwise} \end{array} \right. \mbox{,}
\end{equation}
\begin{equation}
 G_{2i,2j-1} = \left\{ \begin{array}{cl} \frac{j(N-i)}{4(N+1)} & \hspace{0.2in}
 \mbox{if~} i+1 > j \vspace{0.08in} \\ \frac{(i+1)(N+1-j)}{4(N+1)} &
 \hspace{0.2in} \mbox{otherwise} \end{array} \right. \mbox{,}
\end{equation}
\begin{equation}
 G_{2i-1,2j} = \left\{ \begin{array}{cl} \frac{(i-1)(N+1-j)}{2(N+1)} &
 \hspace{0.2in} \mbox{if~} i-1 < j \vspace{0.08in} \\ \frac{j(N+2-i)}{2
 (N+1)} & \hspace{0.2in} \mbox{otherwise} \end{array} \right. \mbox{,}
\end{equation}
 and
\begin{equation}
 G_{2i-1,2j-1} = \left\{ \begin{array}{cl} \frac{j(N+2-i)}{2(N+1)} &
 \hspace{0.2in} \mbox{if~} i \geq j \vspace{0.08in} \\ \frac{(i-1)(N+1-j)}{2
 (N+1)} & \hspace{0.2in} \mbox{otherwise} \end{array} \right. \mbox{,}
\end{equation}
\label{E:Delta2_Gij}
\end{mathletters}
for $i,j = 1,2,\ldots ,N$. Obviously, the two point correlation function
 $G_{ij}$ varies linearly with the distance between sites $i$ and $j$. Upon a
 uniform and random particle addition, $1/f^2$ scaling is observed as
 $N\!\rightarrow\!\infty$ and the model is indeed self-organized critical (but
 with a trivial exponent).
\par
 This model is not equivalent to any ``ordinary'' one-dimensional ASM with
 toppling rules being translational invariant except possibly at the boundary.
 Suppose the contrary, we can find a set of $h_i^{new} = f_i ( h_1,\ldots
 ,h_{2N})$ such that the system becomes a one-dimensional ASM after applying
 these transformations. $f_i$, however, is independent of $h_j$ for all $j \neq
 i$ or else toppling of site $i$ must depend on the neighboring sites, making
 the transformed system not an ASM. It is easy to check that any transformation
 $f_i = f_i (h_i)$ cannot reduce Eq.~(\ref{E:Delta2_Def}) to a Toeplitz form
 and hence this model is not equivalent to any ``ordinary'' one-dimensional
 ASM.
\section{Conclusions}
 In summary, we have explicitly constructed the simplest possible class of
 cellular automaton examples exhibiting self-organized criticality without the
 presence of local particle conservation law: namely, Abelian Sandpile Models
 on one-dimensional superlattices. Moreover, a simple method of finding
 recurrence phase space configurations, based on the idea of the burning
 algorithm is introduced, which is useful when the dimension of the system is
 low.
\appendix
\section*{Appendix: Finding $\Delta^{-1}$}
 We rearrange the site labels of the system using the map:
\begin{equation}
 \left\{ \begin{array}{ccc} 2i & \longrightarrow & i \\ 2i-1 & \longrightarrow
 & N+i \end{array} \right.
\end{equation}
 for $i=1,2,\ldots ,N$, the toppling matrix $\Delta$ can be re-written as
\begin{equation}
 \Delta_{new} \equiv \Delta^{(2N)}_{new} = \left[ \begin{array}{ll} 2D^T & B \\
 2B^T & D \end{array} \right]
\end{equation}
 where $B^T$ denotes the transpose of $B$. Moreover, $B$ and $D$ are
 $N\!\times\!N$-matrices whose elements are given by
\begin{mathletters}
\begin{equation}
 B_{ij} = \left\{ \begin{array}{rl} -1 & \hspace{0.2in} \mbox{if~} j-i = 1 \\
 0 & \hspace{0.2in} \mbox{otherwise} \end{array} \right.
\end{equation}
 and
\begin{equation}
 D_{ij} = \left\{ \begin{array}{rl} 2 & \hspace{0.2in} \mbox{if~} i = j \\ -1 &
 \hspace{0.2in} \mbox{if~} i-j = 1 \\ 0 & \hspace{0.2in} \mbox{otherwise}
 \end{array} \right.
\end{equation}
\end{mathletters}
respectively. According to the block matrix inversion formula,
\begin{equation}
 \Delta_{new}^{-1} = \left[ \begin{array}{cc} 0.5\ X^{-1} & -0.5 \left( D^T
 \right)^{-1} B\ Y^{-1} \\ -D^{-1} B^T X^{-1} & Y^{-1} \end{array} \right]
 \label{E:NewDeltaInv}
\end{equation}
 where $X = D^T - B\ D^{-1} B^T$ and $Y = D - B^T \left( D^T \right)^{-1} B$.
 Since $D$ is a Toeplitz matrix, $D^{-1}$ can be evaluated easily
 \cite{Toeplitz} and is given by
\begin{equation}
 D^{-1}_{ij} = \left\{ \begin{array}{ll} 0.5^{i-j+1} & \hspace{0.2in}
 \mbox{if~} i \geq j \\ 0 & \hspace{0.25in} \mbox{otherwise} \end{array}
 \right. \mbox{.} \label{E:DInv}
\end{equation}
 As a result,
\begin{equation}
 X_{ij} = \left\{ \begin{array}{ll} 2 & \hspace{0.2in} \mbox{if~} i = j = N \\
 -1 & \hspace{0.2in} \mbox{if~} i-j = 1 \\ 1.5 & \hspace{0.2in} \mbox{if~} i =
 j \mbox{~and~} i < N \\ -0.5^{i-j+1} & \hspace{0.2in} \mbox{if~} i > j
 \mbox{~and~} i < N \\ 0 & \hspace{0.2in} \mbox{otherwise} \end{array} \right.
 \mbox{.}
\end{equation}
 We divide $X$ into four block matrices by partitioning the sites into two
 sets, namely: $\{ 1,2,\ldots ,N-1 \}$ and $\{ N \}$. The only
 $(N\!-\!1)\!\times\!(N\!-\!1)$-matrix so formed is in Toeplitz form whose
 inverse can be found readily \cite{Toeplitz}. After that, by means of the
 block matrix inversion formula again, we obtain
\begin{equation}
 X^{-1}_{ij} = \left\{ \begin{array}{cl} \frac{j(N-i)}{2(N+1)} & \hspace{0.2in}
 \mbox{if~} i > j \vspace{0.08in} \\ \frac{(i+1)(N+1-j)}{2(N+1)} &
 \hspace{0.2in} \mbox{otherwise} \end{array} \right. \mbox{.} \label{E:XInv}
\end{equation}
\par \indent
 Using the same method, we find that
\begin{equation}
 Y^{-1}_{ij} = \left\{ \begin{array}{cl} \frac{j(N+2-i)}{2(N+1)} &
 \hspace{0.2in} \mbox{if~} i \geq j \vspace{0.08in} \\ \frac{(i-1)(N+1-j)}{2
 (N+1)} & \hspace{0.2in} \mbox{otherwise} \end{array} \right. \mbox{.}
 \label{E:YInv}
\end{equation}
 Now $D^{-1} B^T X^{-1} $ and $\left( D^T \right)^{-1} B\ Y^{-1}$ can be
 evaluated using Eqs.~(\ref{E:DInv}),~(\ref{E:XInv}) and~(\ref{E:YInv}). Thus
 all four block matrices in Eq.~(\ref{E:NewDeltaInv}) are computed. Finally,
 Eqs.~(\ref{E:Delta2_Gij}a)--(\ref{E:Delta2_Gij}d) are obtained by changing
 the labels of the sites back to the original ones.
\acknowledgments{I would like to thank the Aspen Center for Physics for its
 hospitality during which part of the work was done. This work is supported by
 DOE grant DE-FG02-90ER40542.}


\begin{references}
\bibitem{BTW} P. Bak, C. Tang, \& K. Wesienfeld, {\it \prl}, {\bf 59}, 381
 (1987); P. Bak, C. Tang, \& K. Wesienfeld, {\it \pra}, {\bf 38}, 364 (1988).
\bibitem{ASM} D. Dhar, {\it \prl}, {\bf 64}, 1613 (1990).
\bibitem{Hwa} T. Hwa, \& M. Kardar, {\it \prl}, {\bf 62}, 1813 (1989).
\bibitem{Spec} G. Grinstein, D. H. Lee, \& S. Sachdev, {\it \prl}, {\bf 64},
 1927 (1990); S. S. Manna, L. B. Kiss, \& J. Kert\'{e}sz, {\it J.\ Stat.\
 Phys.\ }, {\bf 61}, 923 (1990); K. Christensen, Z. Olami, \& P. Bak, {\it
 \prl}, {\bf 68}, 2417 (1992); J. E. Socolar, G. Grinstein, \& C. Jayaprakash,
 {\it \pre}, {\bf 47}, 2366 (1993).
\bibitem{FFM} K. Chen, P. Bak, \& M. H. Jensen, {\it Phys.\ Lett.\ A\ }, {\bf
 149}, 207 (1990); B. Drossel, \& F. Schwabl, {\it \prl}, {\bf 69}, 1629
 (1992); S. Clar, B. Drossel, \& F. Schwabl, {\it \pre}, {\bf 50}, 1009
 (1994).
\bibitem{Other} P. Ruelle, \& S. Sen, {\it J.\ Phys.\ A\ }, {\bf 25}, L1257
 (1992); A. A. Ali, \& D. Dhar, cond-mat preprint \#9412085 (1994).
\bibitem{BA} D. Dhar, \& S. S. Manna, {\it \pre}, {\bf 49}, 2684 (1994).
\bibitem{Speer} E. R. Speer, {\it J.\ Stat.\ Phys.\ }, {\bf 71}, 61 (1993).
\bibitem{GASM} H. F. Chau, \& K. S. Cheng, {\it J.\ Math.\ Phys.\ }, {\bf 34},
 5109 (1993).
\bibitem{Aval_Finite} S. W. Chan, \& H. F. Chau, submitted (adap-org preprint
 \#9410002, 1994).
\bibitem{Inv_A} H. F. Chau, {\it \pre}, {\bf 50}, 4226 (1994).
\bibitem{Relative_Scale} H. F. Chau, \& C. Ho, {\it \pre}, {\bf 49}, 902
 (1994).
\bibitem{Toeplitz} G. Heinig, \& K. Rost, {\it Algebraic Methods for
 Toeplitz-like Matrices and Operators}, Chap. 1 (Birkh\"{a}user Verlag,
 Stuttgart, 1984).
\end{references}
\end{document}